\documentclass[aps,twocolumn,notitlepage,longbibliography,nofootinbib,superscriptaddress
]{revtex4-2}
\usepackage[utf8]{inputenc}
\usepackage[
bookmarks=true,bookmarksnumbered=true,
colorlinks = true,anchorcolor= blue, 
linkcolor = darkred, citecolor = darkgreen
]{hyperref}
\usepackage{ulem}
\usepackage{amsfonts,amsthm,amsmath,amssymb,graphicx}
\usepackage{braket}
\usepackage{bm}
\usepackage{float}
\usepackage{mathrsfs}
\usepackage{booktabs}
\usepackage{makeidx}
\usepackage{color}
\usepackage{tikz}
\usepackage{here}
\numberwithin{equation}{section}
\usepackage{color}
    \definecolor{darkgreen}{rgb}{0,0.5,0}
     \definecolor{darkblue}{rgb}{0,0,0.6}
    \definecolor{purple}{rgb}{0.4,0.2,0.7}
    \definecolor{darkred}{rgb}{0.7,0,0}
    \definecolor{airforceblue}{rgb}{0.36, 0.54, 0.66}
    \definecolor{cyan}{rgb}{0.0, 1.0, 1.0}
  	\definecolor{cyan(process)}{rgb}{0.0, 0.72, 0.92}

\usepackage{comment}

\usepackage{comment}

\newcommand{\be}{\begin{equation}}
\newcommand{\ee}{\end{equation}}
\newcommand{\ba}{\begin{eqnarray}}
\newcommand{\ea}{\end{eqnarray}}
\newcommand{\f}{\frac}
\newcommand{\s}{\sqrt}

\def\ba#1\ea{\begin{align}#1\end{align}}

\def\f {\frac}

\begin{document}
\title{Bridging two quantum quench problems -- 
	local joining quantum quench
	and M\"obius quench --
	and their holographic dual descriptions
}
\author{Jonah Kudler-Flam }
\email{jkudlerflam@ias.edu}
\affiliation{School of Natural Sciences, Institute for Advanced Study, Princeton, New Jersey, 08540, USA}
\affiliation{Princeton Center for Theoretical Science, Princeton University, Princeton, New Jersey, 08544, USA}
\author{Masahiro Nozaki}
\email{masahiro.nozaki@riken.jp}
\affiliation{Kavli Institute for Theoretical Sciences, University of Chinese Academy of Sciences,
Beijing 100190, China}
\affiliation{RIKEN Interdisciplinary Theoretical and Mathematical Sciences (iTHEMS),
Wako, Saitama 351-0198, Japan}
\author{Tokiro Numasawa }
\email{numasawa@issp.u-tokyo.ac.jp} 
\affiliation{Institute for Solid State Physics, University of Tokyo, Kashiwa 277-8581, Japan}
\author{Shinsei Ryu }
\email{shinseir@princeton.edu}
\affiliation{Department of  Physics, Princeton University, Princeton, NJ 08544}
\author{Mao Tian Tan}
\email{maotian.tan@apctp.org}
\affiliation{Asia Pacific Center for Theoretical Physics, Pohang, Gyeongbuk, 37673, Korea}

\begin{abstract}
We establish an equivalence between two different quantum quench problems, the joining local quantum quench and the M\"obius quench,
in the context of $(1+1)$-dimensional conformal field theory (CFT).
Here, in the former, two initially decoupled systems (CFTs) on finite
intervals are joined at $t=0$.
In the latter, we consider 
the system that is initially prepared in the ground state of
the regular homogeneous Hamiltonian on a finite interval 
and, after $t=0$, let it time-evolve by the 
so-called M\"obius Hamiltonian
that is spatially inhomogeneous. 
The equivalence allows us to relate the time-dependent physical observables
in one of these problems to those in the other.
As an application of the equivalence, we construct a holographic dual
of the M\"obius quench from that of the local quantum quench.
The holographic geometry involves an end-of-the-world brane whose profile exhibits non-trivial dynamics. 

\end{abstract}

\maketitle

{\hypersetup{linkcolor= red, filecolor = magenta, urlcolor=magenta}}

\section{Introduction}

Spatial inhomogeneity is ubiquitous in quantum many-body problems
and can lead to a rich variety of physics.
On the one hand, it can take the form of randomness or disorder.
An important phenomenon caused by disorder is the Anderson localization \cite{PhysRev.109.1492}, which plays an important role in the integer quantum hall effect for example.
On the other hand, it can also be introduced in
a more controlled manner, such as a harmonic trap of cold atomic gas \cite{2008RvMP...80..885B}.
For example, quantum field theory 
can be studied on a curved spacetime \cite{SciPostPhys.2.1.002}, a setting that appears in many contexts of physics \cite{Birrell:1982ix}.

In this paper, we are interested in a particular kind of
spatial inhomogeneity that is introduced
to many-body quantum systems in one spatial dimension.
It can be obtained from the regular,
homogeneous Hamiltonian $H_0$
--
given as a spatial integral of the Hamiltonian density $h(x)$
as $H_0 = \int dx\, h(x)$
--
by deforming it by introducing an envelop function $f(x)$,
$H = \int dx\, f(x)\, h(x)$.
In particular, we will be interested in the so-called M\"obius deformation and sine-square deformation (SSD).
(See Sec.\ \ref{sec:MobiusQ} for the choice of the envelope function and more details.) 
One of the initial motivations for these deformations
was to study many-body systems with open boundaries
numerically while suppressing the boundary effects \cite{2009PThPh.122..953G}.
Amazingly, at a conformal quantum critical point,
an SSD Hamiltonian has exactly the same ground state as
that of the regular Hamiltonian with periodic boundary condition
\cite{2011PhRvB..83f0414H, 2012JPhA...45k5003K,2015MPLA...3050092T}.
More recently, the M\"obius deformation and SSD
have been used to study non-equilibrium dynamics \cite{2018PhRvB..97r4309W,2021arXiv211214388G,2018arXiv180500031W,2021PhRvB.103v4303L}.
Inhomogeneities  in CFT are also studied in
\cite{2016JSMTE..05.3108A, SciPostPhys.2.1.002, 2017ScPP....3...19D,2018JSP...172..353G,2019PhRvL.122b0201L,
MacCormack:2018rwq} for example.

The spatial deformation of the above kind
appears in various contexts.
Another example is the modular Hamiltonian 
(also known as the entanglement Hamiltonian),
defined for a reduced density matrix for a subregion,
is given in some cases by a spatial deformation of the regular Hamiltonian (in the above sense).
The modular Hamiltonian for the ground state (vacuum) of a relativistic invariant theory,
when half of the total space is traced out, is nothing but the Rindler Hamiltonian,
and the evolution by a spatially deformed Hamiltonian appears in that context \cite{Bisognano:1975ih,Bisognano:1976za}.
In high-energy physics, understanding the evolution by modular Hamiltonians
is important to study the structure of spacetime through the AdS/CFT correspondence
\cite{Maldacena:1997re, Jafferis:2015del,Czech:2012bh,Dong:2016eik, Faulkner:2017vdd, Faulkner:2018faa,Chen:2018rgz}.

Although the action of these spatially deformed Hamiltonians on special states is understood through the relation
to the undeformed Hamiltonians, the properties of general excited states
are yet to be understood.
To develop a deeper understanding,
in this paper, we will consider two seemingly different quantum quench problems in
the context of (1+1)-dimensional conformal field theory (CFT).
First, we consider a quantum quench process,
which we call the M\"obius quench,
where the system is initially prepared for the ground state of CFT 
(with the regular Hamiltonian) on a finite interval. 
At $t=0$, the system's Hamiltonian is changed from the regular Hamiltonian to the M\"obius Hamiltonian.
This quench problem was studied in 
Ref.\ \cite{2018PhRvB..97r4309W}.
Quench problems with more general spatial deformations are studied in \cite{Princetonpreprint}.
In the second quench problem, we initially consider two decoupled systems (ground states of CFT), 
each defined on a finite interval of equal length. 
The two systems are joined or ``glued'' at $t=0$ and
then time-evolved by the 
uniform CFT Hamiltonian of the coupled intervals
\cite{2011JSMTE..08..019S}.
We call this the local quantum quench.
Quantum quenches in CFT, including local quantum quench,  were studied in various context \cite{2006PhRvL..96m6801C, 2007JSMTE..06....8C, 2007JSMTE..10....4C, 2016JSMTE..06.4003C, 2012JPhA...45J2001B, 2015AnHP...16..113B,2016JSMTE..06.4005B,Wen:2016bzx}.

One of the main results of the paper is to 
establish the equivalence between these two problems.
Not only does the equivalence allow us to relate the time-dependent physical observables,
but also to gain a deeper understanding of aspects of these quantum quench problems. 
Here, we note that in (1+1)d CFT 
many non-equilibrium
problems are related to each other by conformal mappings.
In particular, 
all quantum quench problems for which 
the relevant spacetime geometry can be mapped to the 
upper half-plane are related to each other.
These include, e.g., 
inhomogeneous global quenches, 
finite-size global quenches, splitting local quenches, double
local quenches, Floquet CFT, etc.
They only differ by the space-dependent Weyl transformation and coordinate transformation. 
Especially the Weyl transformation doesn't affect the time evolution.
For example,
physical observables in these quench problems exhibit 
eternal oscillations, albeit the CFT in question can
be a fast quantum information scrambler (in the limit of large central charge).
The oscillations can be attributed to the
underlying ${\it SL}(2, \mathbb{R})$ structure
of the M\"obius Hamiltonian  
\cite{2018PhRvB..97r4309W}.
The non-trivial mapping between the two problems
also allows us to construct their holographic dual (AdS/CFT) descriptions easily. 
We find that in the holographic dual descriptions,
the so-called end-of-the-world (EOW) brane is involved in the bulk \cite{Takayanagi:2011zk,Fujita:2011fp},
the dynamics of which describes the time-dependence of 
physical observables (entanglement entropy, energy density). 
Here, we note that the EOW brane
is a key ingredient of holographic duality for boundary CFT (BCFT). 
We will also speculate that similar equivalence relations
can be established for a wider class of quantum quench problems.

The rest of this paper is organized as follows.
In Sec.\ \ref{sec:localQ}, we review
the local quench on a finite strip and study
the entanglement and energy-density dynamics.
In Sec.\ \ref{sec:MobiusQ},
we study the M\"obius quench and discover its relation to
the local quench problem.
Using this relation, we also study the energy-momentum tensor dynamics
in the M\"obius quench.
In Sec.\ \ref{sec:holography},
we construct the holographic dual of the M\"obius quench
using the relation between the two quench problems.
In particular, we study the end-of-the-world brane dynamics
and compare it with the entanglement dynamics in CFT analysis.
We conclude in Sec.\ \ref{sec:conclusion},
and provide some future discussions.

\section{A warm up: Global quench and Rindler Quench}
First, we consider global quenches on an infinite line $(-\infty, \infty)$.
We imagine that first we have a gapped deformation of conformal field theory and then suddenly turn off the deformation term.
The initial gapped ground state is evolved by the homogeneous CFT Hamiltonian.
To approximate the gapped ground state, we use the smeared boundary state \cite{2005JSMTE..04..010C,2013JHEP...05..014H,2016JSMTE..06.4003C}
\be
\ket{\psi_0} \propto e^{-\f{\beta}{4}H} \ket{B}.
\ee
The evolution of entanglement entropy on a half line $[0,\infty)$ becomes \footnote{Here we simply omit a non-universal term, which is denoted as $\tilde{c}_1'$ in \cite{2013JHEP...05..014H}. 
}
\be
S_A = \f{c}{6} \log \Big( \f{\beta}{\pi z_{\epsilon}} \cosh \big(\f{2\pi t}{\beta} \big)\Big)
\ee
Here $z_{\epsilon}$ is a UV cutoff.
Note that entanglement entropy for even an infinite line is well-defined reflecting the fact that the initial state only has short-range entanglement.
Entanglement entropy for an infinite line grows linearly in time forever.
When we consider a finite interval instead, entanglement entropy saturates when the time reaches the half of the length of the interval divided by the speed of sounds.

Next, we consider the Rindler quenches.
In this problem, we start from the ground state $\ket{G}$ of the homogeneous Hamiltonian $H_0 = \int_0^{\infty} h(x) dx  $ on a half line $[0,\infty)$.
Then, we change the Hamiltonian to the Rindler Hamiltonian $H_1 = a \int_0^{\infty} x h(x) dx $ and evolve the original state $\ket{G}$ by the Rindler Hamiltonian.
Here $a$ is a parameter of the dimension of the inverse of the length.
This is equivalent to putting CFT on a curved spacetime 
\be
ds^2 = - a^2 x^2 dt^2 + dx^2. \label{eq:RindlerMetric}
\ee
Then, the evolution of entanglement entropy for an infinite line $[x,\infty)$ is given by 
\be
S_A = \f{c}{6} \log \Big(\f{2x}{\epsilon} \cosh \big(a t\big) \Big)
\ee

Two quench problems show a similar evolution of entanglement entropy.
Actually, after identifying the parameter $a  = \f{2\pi }{\beta}$ and changing the cutoff $z_{\epsilon} = \f{\epsilon}{a x}$, the entropy for a global quench becomes 
\be
\f{c}{6} \log \Big(\f{\beta}{\pi z_{\epsilon} }  \cosh \big(\f{2\pi t}{\beta} \big) \Big) = \f{c}{6} \log \Big(\f{2 x}{\epsilon }  \cosh(at) \Big), \label{eq:EERindlerfromGlobql}
\ee
and we can obtain exactly the same evolution of the entropy for Rindler quenches.

Actually, these two problems are related in a more direct manner.
First, the Euclidean version of the metric \eqref{eq:RindlerMetric} is 
\ba
ds^2 &= a^2 x^2 d\tau^2 + dx^2 \notag \\
&= d \tau_P^2 + dx_P^2, \label{eq:RindlerToPoincare}
\ea
where we used the coordinate transformation 
\be
\tau_P = x \sin (a \tau ), \quad x_P = x \cos (a \tau )
\ee
The coordinate transformation suggests that the ground state of the homogeneous Hamiltonian is equivalent to the boundary state with the finite amount of Euclidean evolution by the Rindler Hamiltonian:
\be
\ket{G} = e^{-\f{\pi}{2a} H_{1}}\ket{B}.
\ee
Next, changing the spacial coordinate $x = e^{a \rho}/a$, we obtain
\ba
ds^2 &= a^2 x^2 d\tau^2 + dx^2 \notag \\
&= ( e^{a\rho})^2 (d\tau^2 + d\rho^2), \label{eq:RindlerToCC}
\ea
which means that after Weyl transformation $ds^2 \to  e^{-2a\rho}ds^2$, the Euclidean path integral is equivalent to that of Calabrese-Cardy state preparation for global quenches \cite{2005JSMTE..04..010C,2013JHEP...05..014H,2016JSMTE..06.4003C}.
These two show that the correlation functions after Rindler quench is Weyl equivalent to those after global quenches:
\ba
&\braket{\mathcal{O}_1(t_1,x_1) \cdots\mathcal{O}_n(t_n,x_n)}_{\text{Rindler}} \notag \\
= & e^{-\Delta_1 a\rho_1} \cdots e^{-\Delta_n a \rho_n} \braket{\mathcal{O}_1(t_1,x_1) \cdots\mathcal{O}_n(t_n,x_n)}_{\text{Global}}.
\ea 
In particular, we can apply this relation to twist operators to study entanglement entropy and we can deduce the relation \eqref{eq:EERindlerfromGlobql}.
In this manner, we can explain the relation between Rindler quench and global quench following their path integral representations and coordinate transformations among them.
\begin{figure}[t]
\begin{center}
\includegraphics[width=8cm]{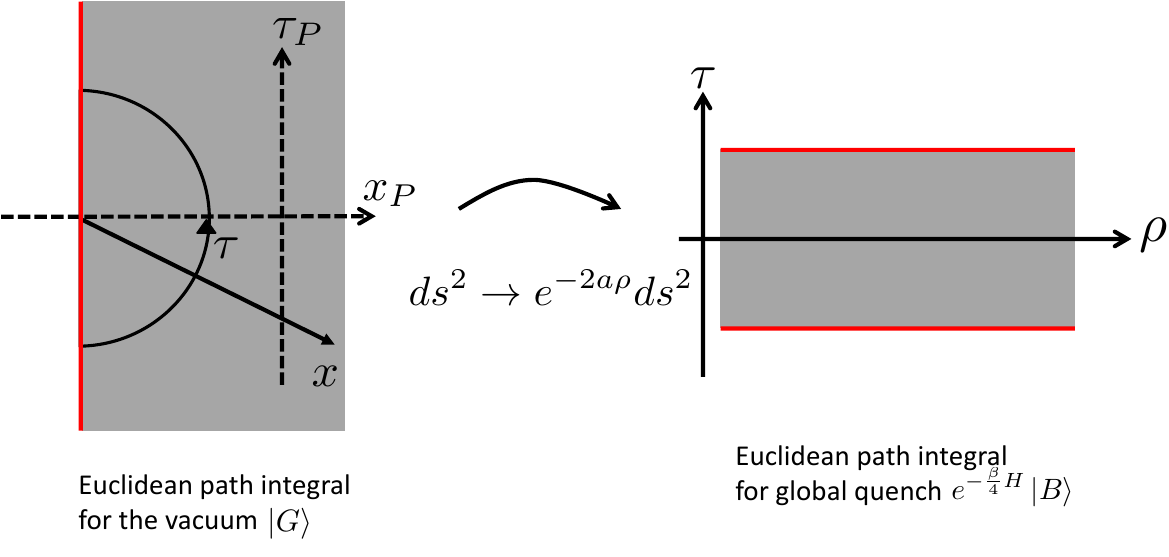}
\end{center}
\caption{
The map from the Euclidean integral for the vacuum state $\ket{G}$ to that for the globally-quenched state $e^{-\f{\beta}{4}H}\ket{B}$. The left and right panels illustrate the Euclidean path integrals for $\ket{G}$ and $e^{-\f{\beta}{4}H}\ket{B}$, respectively.
}
\label{fig:RindlerCoordinates}
\end{figure}

\section{Local quench on finite strips } 
\label{sec:localQ}

In Ref.\ \cite{2011JSMTE..08..019S},
the authors studied a local quantum quench process
in the context of (1+1)d CFT.
In this process, the system is initially ``cut'' into two independent subsystems.
To be specific, we consider two intervals of equal length ($=L_{{\it eff}}/2)$,
$[-L_{{\it eff}}/2, 0]$, and $[0, L_{{\it eff}}/2]$.
(Here, denoting the total system size by $L_{{\it eff}}$ may look bizarre.
The motivation for this will become clear when we later make 
contact with the M\"obius quench.)
The system is initially prepared as the tensor product of  
the ground states of the two intervals.
These two intervals are then glued together at time $t=0$. 
Namely, for $t>0$, the system time-evolves
by the Hamiltonian for the single interval of length $L_{{\it eff}}$.

\begin{figure}[t]
\begin{center}
\includegraphics[width=8cm]{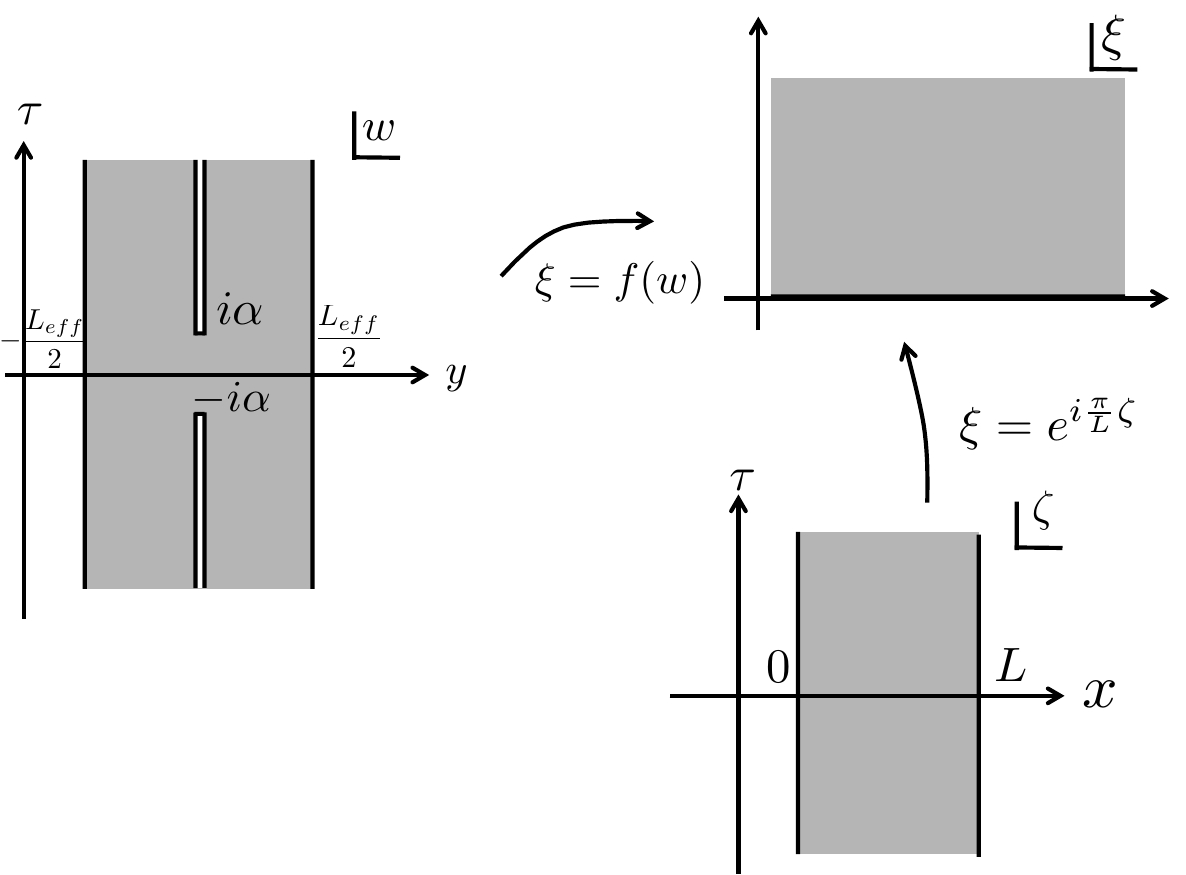}
\end{center}
\caption{The Euclidean geometry for the local quench on a finite interval.
The left figure is the geometry to represent the joining of two intervals with the regularization parameter $\alpha$ whereas the right figure is the upper half plane after the conformal transformation \eqref{eq:LocalQfinMap}.
}
\label{fig:FiniteQmap}
\end{figure}

The quench process can be analyzed by using the
Euclidean path integral on a ``pants'' geometry,
which is represented in Fig.\ \ref{fig:FiniteQmap}.
We use $w=y+ i \tau$ to coordinatize this geometry
where $\tau$ and $y$ represent Euclidean temporal and spatial 
coordinates, respectively.
We regularize this excited state by the Euclidean path integral
for Euclidean time $\alpha$.

The Euclidean geometry is mapped to the upper half plane by the conformal transformation \cite{2011JSMTE..08..019S}
\be
\label{eq:LocalQfinMap}
\xi = f(w) =   i \s{\f{\sin (\f{\pi}{L_{{\it eff}}} (i\alpha + w))}{\sin
(\f{\pi}{L_{{\it eff}}} (i\alpha - w))}}
.
\ee
Physical observables can then be computed from the corresponding
correlations on the upper half-plane.
For example, the one -oint function
of a primary operator $O(w)$ with conformal dimension
$\Delta$ 
$
\braket{O(w,\bar{w})} = A_O
|({1}/{2 \text{Im}\, \xi}) ({d \xi}/{d w}) |^{\Delta}
$ where $O(w,\bar{w})$ is a primary operator, $\Delta = 2h$ is the scaling dimension of $O$ with the conformal weight $h$ and $A_{O}$ is the one point function. 
Introducing the mapping to the strip $\xi = e^{i \f{\pi}{L}\zeta}$, which will use later, we can also write the map as 
\begin{align}
\label{eq:LocalQfinMap2}                        
e^{i\f{\pi}{L}\zeta}                            
= i\s{\f{ i \tanh \f{\pi \alpha}{L_{{\it eff}}} 
+ \tan \f{\pi w}{L_{{\it eff}}}}                
{ i \tanh \f{\pi \alpha}{L_{{\it eff}}}         
- \tan \f{\pi w}{L_{{\it eff}}}}}.              
\end{align}

\begin{figure}[t]
\begin{center}
\includegraphics[width=6cm]{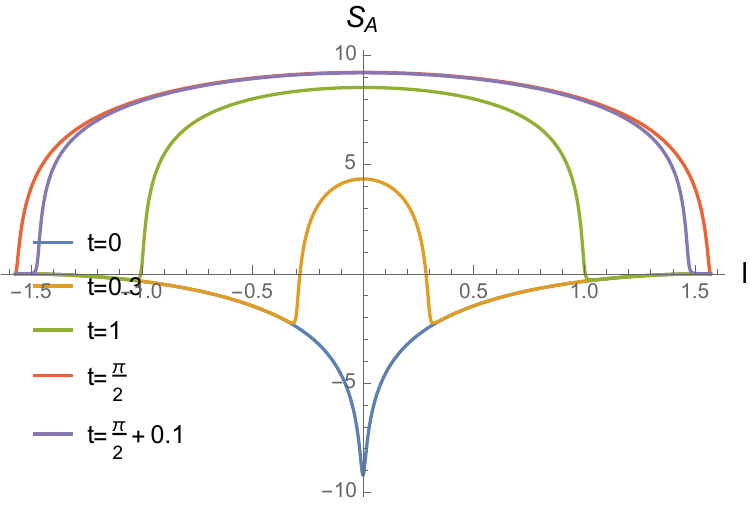}
\end{center}
\caption{
The time-dependence of the entanglement entropy
after the joining quantum quench, \eqref{eq:FiniteLocalQent}. 
Here, we subtract the ground state entanglement entropy
of the interval $[-L_{{\it eff}}/2, L_{{\it eff}}/2]$.
Here we set $L_{{\it eff}} = \pi$.
}
\label{fig:EOWprofileEnt}
\end{figure}

The entanglement entropy can be obtained from
the correlation function of the twist-anti-twist operators
\cite{2004JSMTE..06..002C}.
The Euclidean time $\tau$ is analytically continued to
Lorentzian time $t$, $\tau \to i t$.
For general $t$, and for the subsystem of an interval
$y\in [-L_{{\it eff}}/2, +l/2]$,
we obtain the following expression for the entanglement entropy:
\begin{align}
S_A(t,l) & = \f{c}{12} \log \bigg[ 
\left( \f{2L_{{\it eff}}}{\pi z_\epsilon} \right) ^2
\f{1}{ 2\sinh^2 \f{2\pi \alpha}{L_{{\it eff}}}}
\notag \\
& \quad \qquad            
\times
\Big(M(t,l)^2 +
M(t,l)N(t,l) \Big) \bigg],
\label{eq:FiniteLocalQent}
\end{align}
where $c$ is the central charge, $z_\epsilon$ is a UV cutoff,
and 
\ba
M(t,l) &= \s{N(t,l)^2 + \sin^2 \f{2\pi l}{L_{{\it eff}}} \sinh^2 \f{2\pi \alpha}{L_{{\it eff}}}},\notag \\
N(t,l) &= \cos \f{2\pi l}{L_{{\it eff}}} \cosh \f{2\pi \alpha}{L_{{\it eff}}} - \cos \f{2\pi t}{L_{{\it eff}}}.
\ea
In particular, at $t = 0$, the entanglement entropy just after joining is 
\begin{align}                                                                                 
S_A =  \f{c}{6} \log                                                                                                                                                                     
\left(                                                                                                                                                                                   
\f{2L_{{\it eff}}}{\pi z_\epsilon} \f{\cos \f{\pi l}{L_{{\it eff}} }}{\cosh\f{\pi \alpha}{L_{{\it eff}} } } \s{\sinh^2 \f{\pi \alpha}{L_{{\it eff}} } +\sin^2 \f{\pi l}{L_{{\it eff}} }} 
\right).                                                                                                                                                                                 
\end{align}
Here $z_{\epsilon}$ is a UV cutoff. 
The profile of the dynamical entanglement entropy is plotted in
Fig.\ \ref{fig:EOWprofileEnt}.
Here, we consider the difference between \eqref{eq:FiniteLocalQent} and the ground state entanglement entropy, 
$
S_{{\it diff}} = S_A(t,l) - S_A^{{\it ground}}(l),
$
where
$S_A^{{\it ground}}(l) = \f{c}{6} \log (\f{2 L_{{\it eff}}}
{\pi z_{\epsilon}} \cos \f{\pi l}{L_{{\it eff}}} ) $
is the entanglement entropy of the ground state on the same strip.

\begin{figure}[t]
\begin{center}
\includegraphics[width=6cm]{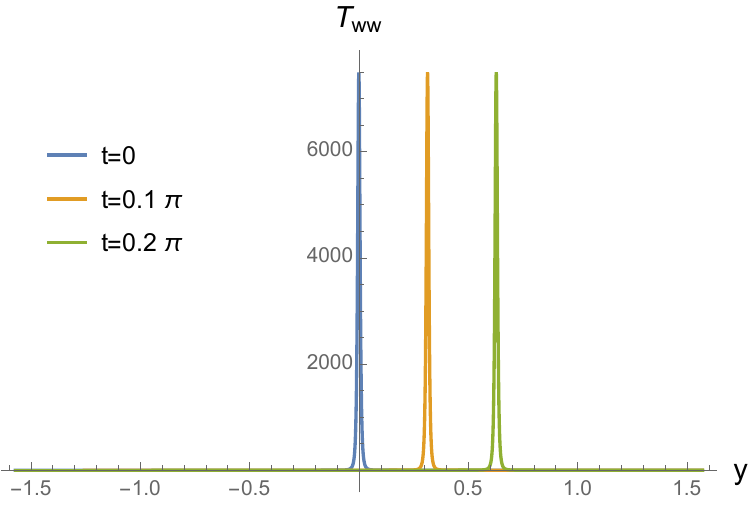} 
\end{center}
\caption{The plot of the stress tensor for right-moving modes.
We set $L_{{\it eff}} = \pi$ and $\alpha = 0.01$ and $c = 6$.
}
\label{fig:StressTplot}
\end{figure}

\begin{widetext}
The conformal map  \eqref{eq:LocalQfinMap}
to the upper half-plane also allows us to compute the time-dependence of the energy-momentum tensor
by 
\ba\label{eq:SchwarzD}
T_{ww}(w) &= -\f{c}{6} \mathcal{L}(w),\quad \mathcal{L}(w) = \f{3(f'')^2 - 2 f' f'''}{4 (f')^2}, \notag \\
\qquad 
\bar{T}_{\bar{w}\bar{w}}(\bar{w}) &= -\f{c}{6} \bar{\mathcal{L}}(\bar{w}), \quad \bar{\mathcal{L}}(\bar{w}) = \f{3(\bar{f}'')^2 - 2 \bar{f}' \bar{f}'''}{4 (\bar{f}')^2}.
\ea

Explicitly, the holomorphic and anti-holomorphic components
of the energy-momentum tensor are given by
\begin{align}
&\mathcal{L}(w) =    
-\f{\pi^2}{8L_{{\it eff}}^2} \f{11 + \cosh(\f{4\pi \alpha}{L_{{\it eff}}}) - 16 \cosh (\f{2\pi \alpha}{L_{{\it eff}}} ) \cos (\f{2\pi w}{L_{{\it eff}}}) + 4 \cos (\f{4\pi w}{L_{{\it eff}}})}{ (\cosh \f{2\pi \alpha}{L_{{\it eff}}} - \cos \f{2\pi w}{L_{{\it eff}}})^2}, \notag \\
&\bar{\mathcal{L}}(w) =  
-\f{\pi^2}{8L_{{\it eff}}^2} \f{11 + \cosh(\f{4\pi \alpha}{L_{{\it eff}}}) - 16 \cosh (\f{2\pi \alpha}{L_{{\it eff}}})  \cos (\f{2\pi \bar{w}}{L_{{\it eff}}}) + 4 \cos (\f{4\pi \bar{w}}{L_{{\it eff}}})}{ (\cosh \f{2\pi \alpha}{L_{{\it eff}}} - \cos \f{2\pi \bar{w}}{L_{{\it eff}}})^2}.
\label{eq:STlocalQ}
\end{align}

\end{widetext}
In Fig.\ \ref{fig:StressTplot},
we plot the right-moving part of the stress tensor.
Right at the moment of the quench,
the stress tensor is sharply peaked at $y=0$,
and then propagates to the right. 
Once the peaks hit the boundaries at $y=\pm L_{{\it eff}}/2$, they get reflected back.
After that the energy-momentum tensor profile
exhibits an eternal oscillation.
Note that at $t = L_{{\it eff}}/2 $ the peak looks like to jump from one boundary to the other.
This jump actually captures the reflection correctly since the right moving excitation is reflected to the left moving excitation at the boundaries.

Note that the eternal oscillations are universal in any two-dimensional CFTs.
In particular, we will encounter the oscillation even in chaotic CFTs like holographic CFTs.
Similar non-thermalizing behaviors are also found in global quenches with boundaries \cite{Kuns:2014zka, Mandal:2016cdw}.

\section{M\"obius quench} \label{sec:MobiusQ}

In this section, we consider another quantum quench problem, 
the M\"obius quench
\cite{2018PhRvB..97r4309W},
which is seemingly different from
the local quantum quench considered in the previous section.
In the M\"obius quench, we start from the ground state
$|{\it GS}\rangle$ of (1+1)d CFT on a finite interval of length $L$,
$H_0 |\Psi_0\rangle = E_{{\it GS}}|\Psi_0\rangle$.
Here, $H_0$ is the (regular) Hamiltonian of CFT on a finite interval,
and given in terms of the energy density operator as
$
H_0  = \int^L_0 dx\, h(x)
$.
At $t=0$, we suddenly change the Hamiltonian from $H_0$ to the M\"obius Hamiltonian
$H_{\text{M\"obius}} = \int_0^{L} dx f_{\gamma}(x)h(x)$ with 
\begin{equation}f_{\gamma}(x) = 1 - \tanh(2\gamma) \cos \left( \frac{2\pi x}{L}\right).
\end{equation}
Here, $\gamma$ is a  real positive parameter.
As we send $\gamma \to 0$ and $\gamma \to + \infty$,
the M\"obius Hamiltonian reduces to
the regular Hamiltonian $H_0$
and the sine-square deformed (SSD) Hamiltonian,
respectively
\cite{2009PThPh.122..953G,
2011PhRvA..83e2118G,
2011PhRvB..83f0414H,
2011PhRvB..84k5116S,
2011PhRvB..84p5132M,
2011JPhA...44y2001K,
2012JPhA...45k5003K,
2012PhRvB..86d1108H,
PhysRevB.87.115128,
2015MPLA...3050092T,
2015JPhA...48E5402I,
2016IJMPA..3150170I,
Okunishi:2016zat,
2016PhRvB..93w5119W,
2018arXiv180500031W,
2020PhRvX..10c1036F,
2020PhRvB.102t5125H,
Fan:2020orx,
2021PhRvR...3b3044W,
2020PhRvR...2b3085L,
Lapierre_2020_1}.
In \cite{2021arXiv211214388G},
the M\"obius quench starting from
a thermal initial state was studied. 
In holographic theories, 
the M\"obius quench induces 
a non-trivial dynamics  
(time-dependent deformation)
of the black hole horizon. 
As we will demonstrate later, 
the current M\"obius quench 
induces a non-trivial dynamics of
the EOW brane.

The M\"obius Hamiltonian effectively changes  the total system size from $L$ to
$L_{{\it eff}}$ where $L$ and $L_{{\it eff}}$ are related by
\cite{Okunishi:2016zat, 2016PhRvB..93w5119W}
\begin{align}
  L_{{\it eff}}
  =
  L \cosh 2\gamma.
\end{align}
Specifically, there is a conformal transformation that maps
the spacetime (cylinder of circumference $L$)
with $H_{\text{M\"obius}}$ as the Hamiltonian,
to another spacetime (cylinder of circumference $L_{{\it eff}}$)
with $H_{0}$ as the Hamiltonian.
In the limit $\gamma\to +\infty$ (the SSD limit), $L_{{\it eff}}\to +\infty$.

\subsection{The equivalence between the M\"obius quench and the local quench on finite strips}

We now establish the equivalence between the local quantum quench in the previous section and the M\"obius quench.
To this end, we first study the relationship between the flat metric and the M\"obius Hamiltonian.
The time-evolution generated by the M\"obius Hamiltonian 
is spatially inhomogeneous and corresponds to the metric
\begin{align}
  ds^2_{\text{M\"obius}}
  &= -
    f_{\gamma}(x)^2
    dt^2
    + dx^2.
\end{align}
The relation between flat metric and the M\"obius Hamiltonian can be read off from 
\begin{align}
  ds^2_{\text{M\"obius}}
  &=
    f_{\gamma}(x)^2
    \Big( -  dt^2 + \Big(
    \frac{dx}{f_{\gamma}(x)}
    \Big)^2
    \Big) \notag \\
&= e^{2 \phi} ( - dt^2 + d y^2 ), \label{eq:MobiusMetric}
\end{align}
where the Weyl factor $e^{2\phi}$ is given by
\begin{align}
  \label{Weyl}
  &e^{2\phi} =
    f_{\gamma}(x)^2,
\end{align}
and $y$ and $x$ are related by the coordinate transformation
\begin{align}
  \label{cood change}
  e^{i \f{\pi}{L}x} =
   i\s{
  \f{i e^{-2\gamma} - \tan \f{\pi y}{L_{{\it eff}}} }
  {i e^{-2\gamma} + \tan \f{\pi y}{L_{{\it eff}}}}
  }. 
\end{align}
By taking $\gamma \to \infty$ while keeping $y$, we can take the SSD limit.
The coordinate transformation in the SSD limit is then given by
\begin{align}
  e^{i \f{\pi}{L}x} =
   i
   \sqrt{
   \frac{ L + 2 i \pi y}{L - 2 i \pi y}
  }. 
\end{align}

Here, $y$ runs from $-\f{L_{{\it eff}}}{2}$ to
$\f{L_{{\it eff}}}{2}$ whereas $x$ runs from $0$ to $L$. 
The latter relation is written as 
\be
e^{2\gamma} \tan \Big(\f{\pi y}{L_{{\it eff}}} \Big) \tan \Big(\f{\pi x}{L} \Big) = -1.
\ee

\begin{figure}[t]
\includegraphics[width=5.6cm]{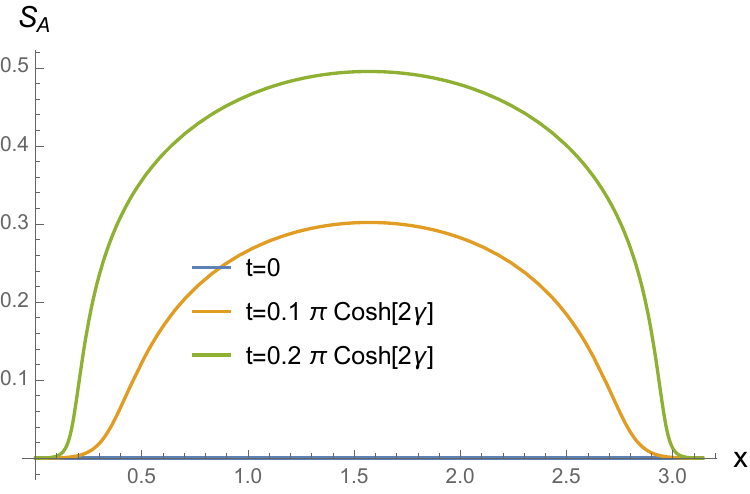}
\caption{ 
The time evolution of entanglement entropy after M\"obius quench  \eqref{eq:EEMobius} with $\gamma = 1$.
For entanglement entropy, we subtract the $t=0$ entropy on an interval $[0, L]$ i.e. $S_A(t)-S_A(0)$. 
}
\label{fig:EEprofile2}
\end{figure}

The coordinate transformation \eqref{cood change}
allows us to relate the flat metric and inhomogeneous metric
corresponding to the M\"obius Hamiltonian. 
In particular, the ground state entanglement entropy of the two problems are related.
Since the M\"obius Hamiltonian shares the same ground
state as the regular Hamiltonian $H_0$,
the entanglement entropy of the ground state (on a finite interval of length $L$) is given by
\begin{align}
  \label{ee Mobius}
  S_A
  &= \f{c}{6} \log
    \left[
     \f{2L}{\pi \epsilon} \sin  \f{\pi x}{L}
    \right],
\end{align} 
where the subsystem $A$ is the interval $[0,x]$.  
By the coordinate change
\eqref{cood change}
and the Weyl transformation of the cutoff, 
\begin{align} \label{eq:cutoff}
	\epsilon \to z_{\epsilon} = 
	\epsilon/f_{\gamma}(x),     
\end{align}
the entanglement entropy \eqref{ee Mobius} becomes 
\ba
S_A &=  \f{c}{6} \log\Big( \f{2L_{{\it eff}}}{\pi z_{\epsilon}} \f{\cos \f{\pi
    y}{L_{{\it eff}} }}{\cosh\f{\pi \alpha}{L_{{\it eff}} } } \s{\sinh^2 \f{\pi
    \alpha}{L_{{\it eff}} } +\sin^2 \f{\pi y}{L_{{\it eff}} }} \Big).
\ea
Here, $\alpha$ satisfies 
\be
 \label{eq:LocalqMobiusqRel}
\cosh \f{2\pi \alpha}{L_{{\it eff}}} = \f{1}{\tanh 2 \gamma},
\ee
or more explicitly $\alpha$ is given as  a function of $\gamma$ by 
\be
\alpha = \f{L}{2\pi} \cosh (2\gamma) 
\text{Arccosh}
\Big(\f{1}{\tanh 2\gamma} \Big). \label{eq:alphagammarel}
\ee
This equation establishes the relation between the parameter $\alpha$ in local quenches, which characterizes the energy scale (and the localization length of the excitation)  of the initial state, and the $\gamma$ that characterizes the inhomogeneity of the M\"obius deformation.
In particular, M\"obius quench with an inhomogeneity parameter $\gamma$ is related to a local quench with the specific parameter $\alpha $ which is determined by \eqref{eq:alphagammarel}.  
The same strategy can be used to establish the relationship between these two problems for $t>0$. 
Once again, by coordinate change and the Weyl transformation
the entanglement entropy for the local quench
\eqref{eq:FiniteLocalQent}
on a finite strip becomes 
\ba
S_A
&=  \f{c}{12} \log \bigg[ \Big( \f{2L^2}{\pi^2 \epsilon^2} \Big)   \Big(f(t,l)^2
+ f(t,l)h(t,l) \Big) \bigg].
\label{eq:EEMobius}
\ea
Here
\ba
h(t,l) &= - \Big(\cos ^2 \f{\pi t}{L_{{\it eff}}} + \cosh 4 \gamma \sin^2
\f{2\pi t}{L_{{\it eff}}} \Big) \cos\f{2\pi l}{L}
\notag \\
& \qquad  +  \sinh 4 \gamma \sin ^2 \f{2\pi t}{L_{{\it eff}}},
\notag \\
f(t,l) &=\s{h(t,l)^2 + \sin ^2 \f{2\pi x}{L}}. 
\ea
This is exactly the time evolution of entanglement entropy after
the M\"obius quench found in \cite{2018PhRvB..97r4309W}.
This suggests that the M\"obius quench is obtained from
the local quench on the $(t,y)$ coordinate
through the Weyl \eqref{Weyl}
and coordinate \eqref{eq:MobiusMetric} transformations.
The evolution of entanglement entropy is shown in Fig.\ \ref{fig:EEprofile2},
where we consider dynamical entanglement entropy with subtraction of the $t=0$
entanglement entropy
$
S^{\text{sub}}(t) = S_A(t) - S_A(0)
$.

Comparing \eqref{eq:LocalQfinMap2} and \eqref{cood change},
if we identify $e^{-2\gamma} = \tanh \f{\pi \alpha}{L_{{\it eff}}}$,
the conformal map for the local quench \eqref{eq:LocalQfinMap2}
gives the holomorphic extension of the spatial coordinate transformation
\eqref{cood change} for the M\"obius quench.
This corresponds to finding the Euclidean path integral representation
of the homogeneous ground state $\ket{\Psi_0}$ using the M\"obius Hamiltonian.
This is somewhat similar to the ground state of $H_0$ on an infinite line
that can also be interpreted as the thermofield double state for the Rindler Hamiltonian.

The conformal symmetry implies that when we consider the sudden quench to the
M\"obius Hamiltonian with the envelop function $\tilde{f}_{\gamma}(x) =
1-\tanh(2\gamma)\cos\f{\pi x}{L}$, 
there is no time evolution.
This can be thought of as a BCFT counterpart of the coincidence of 
the ground states of the M\"obius and the homogeneous Hamiltonians.
On the other hand, in the M\"obius quench we shorten the period of the envelop function $L \to L/2$, which leads to the branch cut structure in \eqref{cood change} and leads to the excitation for the M\"obius Hamiltonian.
Because of the $\mathbb{Z}_2$ symmetry $x \to L -x$ of the envelop function $f_{\gamma}(x)$  and the homogeneous ground state, it is natural to expect that the excitation concentrates near the center $x = L/2$.
What we find is that such a local excitation is represented by the Euclidean path integral \eqref{eq:LocalQfinMap2} for local quenches where the sharp excitation is located near the center but smeared by the regularization $\alpha$.

Mapping the M\"obius quench to the local quench makes it easy to calculate the stress tensor profile.
The stress tensor of the Weyl transformed metric $ds^2 = e^{2\phi}(-dt^2 + dy^2)$ is given by
\be
T_{\mu\nu} = \f{c}{12 \pi} \Big(\hat{T}_{\mu\nu} + T^{\phi}_{\mu\nu} \Big),  \label{eq:StressTMobius}
\ee
where $\hat{T}_{\mu\nu}$ is the stress tensor in the flat metric $ds^2 = -dt^2 + dy^2$ and $T^{\phi}_{\mu\nu}$ is 
\be
T^{\phi}_{\mu\nu} = -  \Big[ \partial _{\mu}\phi\partial_{\nu}\phi - \f{1}{2}\eta_{\mu\nu} \partial^{\rho}\phi\partial _{\rho}\phi - \partial_{\mu}\partial_{\nu}\phi + \eta_{\mu\nu} \partial^{\rho}\partial _{\rho}\phi \Big].
\ee
\begin{widetext}
	In the $(t,x)$ coordinate $T_{\mu\nu}^{\phi}$ becomes  \begin{align}
  &T_{tt}^{\phi} = - \f{2\pi^2 \tanh(2\gamma)}{L^2}
    \Big[ \Big(1 + \cos ^2 \big(\f{2\pi x}{L}\big) \Big) \tanh (2\gamma) -2\cos\big(\f{2\pi x}{L}\big)  \Big],
    \notag \\
  &T_{xx}^{\phi} = - \f{2\pi^2 \sin^2 (\f{2\pi x}{L}) \tanh^2 (2\gamma) }
    {L^2f_{\gamma}(x)^2},
    \notag \\
  &T_{tx}^{\phi} = T_{xt}^{\phi} = 0.
    \label{eq:WeylFactor}
\end{align}
On the other hand, $\hat{T}_{\mu\nu}$ becomes
\begin{align}
	\hat{T}_{tt}(x,t) & = -\f{\pi^2}{8L^2 \cosh^2 2\gamma} \bigg( \f{X(x,t)}{Y(x,t)}+\f{X(x,-t)}{Y(x,-t)}\bigg),                  
	\notag \\
	\hat{T}_{xx}(x,t) & = - \f{\pi^2}{8L^2 f_{\gamma}(x)^2 \cosh^2 2\gamma} \bigg( \f{X(x,t)}{Y(x,t)}+\f{X(x,-t)}{Y(x,-t)}\bigg), 
	\notag \\
	\hat{T}_{tx}(x,t) & = \hat{T}_{xt}(x,t)  \notag                                                                               \\
	                  & = -\f{\pi^2}{8L^2 f_{\gamma}(x) \cosh^2 2\gamma} \bigg( -\f{X(x,t)}{Y(x,t)}+\f{X(x,-t)}{Y(x,-t)}\bigg),   
	\label{eq:EMMobius}
\end{align}  
where $X(x,t)$ and $Y(x,t)$ are given by
\begin{align}
X(x,t) =&  
          10 \sinh^2 2\gamma f_{\gamma}(x)^2
          + 2  \cosh^2 2\gamma
          f_{\gamma}(x)^2
          \notag \\
        & - 16 \sinh 2\gamma
          f_{\gamma}(x)
          \Big\{
           \cosh (2\gamma) g_{\gamma}(x) \cos \f{2\pi t}{L_{eff}} - \sin \f{2\pi x}{L} \sin \f{2\pi t}{L_{eff}}\Big\}  \notag \\
&+ 4 \tanh^2 2\gamma \Big(
 \cosh (2\gamma) g_{\gamma}(x)+  \sin \f{2\pi x}{L}\Big)\Big(
  \cosh (2\gamma) g_{\gamma}(x) -  \sin \f{2\pi x}{L}\Big) \cos \f{4\pi t}{L_{eff}} \notag \\
& - 8  \tanh^2 2\gamma \sin \f{2\pi x}{L}
 \cosh (2\gamma) g_{\gamma}(x) \sin \f{4\pi t}{L_{eff}}
  \\
Y(x,t) =& \bigg[
 \cosh (2\gamma) f_{\gamma}(x)-\sinh (2\gamma) g_{\gamma}(x) \cos \f{2\pi t}{L_{eff}} %
+ \tanh 2\gamma \sin \f{2\pi x}{L} \sin \f{2\pi t}{L_{eff}}\bigg]^2
\end{align}

\end{widetext}
where we defined a function $g_{\gamma}(x) = \tanh 2\gamma - \cos \f{2\pi x}{L}$.
At $t = 0$, the full stress tensor \eqref{eq:StressTMobius} has relatively simple expression as  
\begin{align}
  &T_{tt}(x,0) = -\f{c}{12 \pi }\f{\pi^2}{2 L^2}
    f_{\gamma}(x)^2,
    \notag \\
  &T_{xx}(x,0) =  -\f{c}{12 \pi}\f{4\pi^2}{ L^2}
    \f{\cos \f{2\pi x}{L} \tanh 2\gamma}
    {f_{\gamma}(x)}
    - \f{c}{12 \pi}\f{\pi^2}{2L^2}.
\end{align}

\begin{figure}
\includegraphics[width=5.6cm]{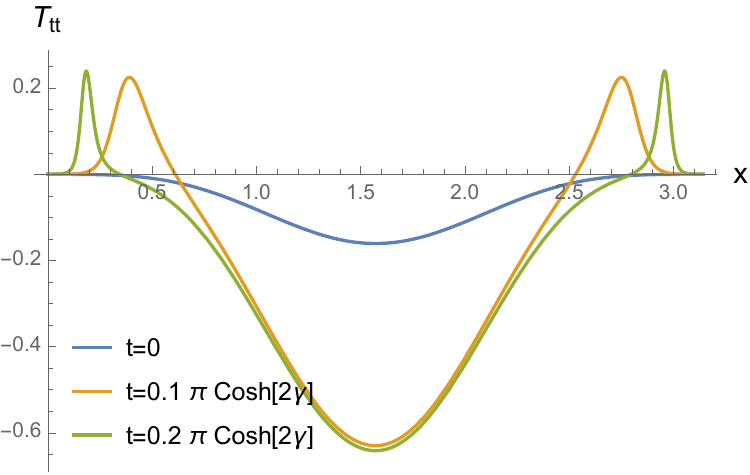}
\caption{ 
  The time evolution of energy density after the M\"obius quench
  \eqref{eq:EEMobius}.
The energy density is given by the $tt$ component of \eqref{eq:StressTMobius}.}
\label{fig:EnergyprofileMob}
\end{figure}

\section{Holographic dual descriptions of quenches} \label{sec:holography}

\subsection{Holographic dual of local quenches on finite strips}

The equivalence we have established allows us to construct the holographic dual description
of one of these quenches starting from that of the other.
Here, we first discuss the holographic dual description of the local quantum quench.
We will later use it to derive the holographic dual of the M\"obius quench.
As the relevant Euclidean path integral is defined on the upper half-plane,
the bulk description is given in terms of AdS/BCFT
\cite{Takayanagi:2011zk,Fujita:2011fp,  2012JHEP...06..066N}.
In AdS/BCFT, what corresponds to BCFT is the bulk AdS space with an end-of-the-world (EOW) brane.

Adopting to our setup, we expect that the EOW is non-stationary in time. 
For the time evolution of states with Euclidean path integral preparation, we can consider the EOW profile in the following manner \cite{Caputa:2019avh}.
We start from CFT defined on the upper half-plane.
The relevant bulk geometry is AdS with the EOW brane
with the metric in $(\eta , \xi, \bar{\xi})$ given by 
\begin{align}
  \label{poincare}
ds^2 = \f{d\eta^2 + d\xi d\bar{\xi}}{\eta^2}.
\end{align}
Assuming the case of tensionless EOW brane for simplicity,
the EOW brane location in $(\eta ,\xi,\bar{\xi})$ is simply given by 
$\xi = \bar{\xi}$.
\begin{figure*}[ht]
\begin{center}
\includegraphics[width=6cm]{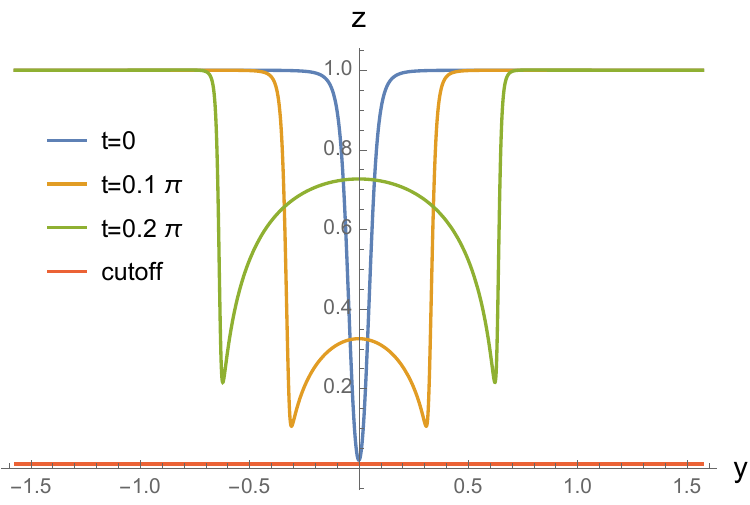}

\hspace{1cm}
\includegraphics[width=6cm]{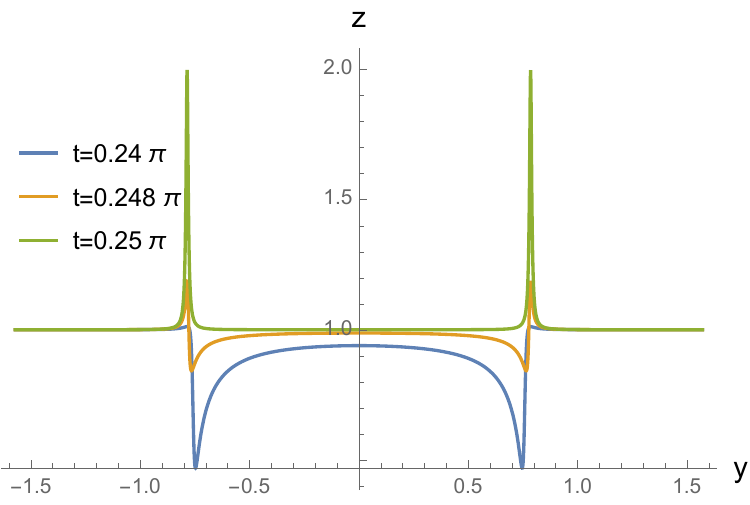}
\end{center}
\caption{
 The holographic dual description of the local quantum quench for finite intervals.
 The EOW brane profile is calculated from \eqref{eq:EOWprofileCoord}.
 We set $L_{{\it eff}} = \pi$.
 {\bf Left:} Early time behavior around $0 < t < 0.2 \f{\pi}{L_{{\it eff}}}$.
 The cutoff is taken to be $z_{\epsilon} = 0.01$.
 {\bf Right:} The behavior near $t = \f{\pi}{4L_{{\it eff}}}$.
 \label{fig:EOWprofile2}
}
\end{figure*}

We now consider the conformal transformation \eqref{eq:LocalQfinMap}
that connects the upper half-plane and the pants geometry.  
In Euclidean signature, the conformal transformation at the boundary 
\be
\xi = f(w), \qquad \bar{\xi} = \bar{f}(\bar{w}),
\ee
is extended to the bulk as \cite{Roberts:2012aq}
\ba
\xi &= f(w) - \f{2z^2 (f')^2 (\bar{f}'')}{4 |f'|^2 + z^2 |f''|^2}, \notag \\
\bar{\xi} &= \bar{f}(\bar{w}) - \f{2z^2 (\bar{f}')^2 (f'')}{4 |f'|^2 + z^2 |f''|^2}, \notag \\
\eta &= \f{4z (f'\bar{f}')^{\f{3}{2}}}{4 |f'|^2 + z^2 |f''|^2}.
\ea
After the coordinate transformation, the metric in $(z,w,\bar{w})$ coordinate is 
\ba
ds^2 &= \f{dz^2 }{z^2} + \mathcal{L}(w)(d\omega)^2 + \bar{\mathcal{L}}(\bar{w})(d\bar{\omega})^2 \notag \\
 & \ \ + \Big(\f{1}{z^2} + z^2 \mathcal{L}(w)\bar{\mathcal{L}}(\bar{w})\Big)dw d\bar{w}, \label{eq:BanadosM}
\ea
which is the general solution of the three-dimensional Einstein gravity \cite{Banados:1998gg}.
Here
\ba
\mathcal{L}(w) &= \f{3(f'')^2 - 2 f' f'''}{4 (f')^2}, \notag \\ 
\bar{\mathcal{L}}(\bar{w}) &= \f{3(\bar{f}'')^2 - 2 \bar{f}' \bar{f}'''}{4 (\bar{f}')^2}.
\ea

In $(z,w,\bar{w})$ coordinate, the EOW brane location is given by $\xi(z,w,\bar{w}) = \bar{\xi}(z,w,\bar{w})$.
Rewriting this condition as $z = z(w,\bar{w})$, we obtain 
\be\label{eq:BraneConfig}
z (w,\bar{w}) = \s{\f{4(f-\bar{f})f'\bar{f}'}{2((f')^2 \bar{f}'' - (\bar{f}')^2 f'') -(f-\bar{f})f''\bar{f}''}}.
\ee

We can apply the above general prescription for the stress tensor and the EOW profile to the conformal map \eqref{eq:LocalQfinMap}.
We already studied the energy-momentum tensor in \eqref{eq:STlocalQ}.
On the other hand, the EOW brane profile is given by
\be
z(y,\tau)^2 = 
 \f{4L_{{\it eff}}^2}{\pi^2}  \f{A(\tau,y)}{B(\tau,y)} \label{eq:EOWprofileCoord},
\ee
where the numerator $A(\tau,y)$ and the denominator $B(\tau,y)$ are given by 
\begin{widetext}
	\begin{align}
		A(\tau,y) 
		  & =  4                                                                                                                       
		\left[ \sin^2 \f{\pi y}{L_{{\it eff}}} \cosh^2 \f{\pi \tau}{L_{{\it eff}}} + \big( \sinh\f{\pi \alpha}{L_{{\it eff}}}  -\cos \f{\pi y}{L_{{\it eff}}} \sinh \f{\pi \tau}{L_{{\it eff}}}  \big) ^2\right]
		\notag
		  &                                                                                                                            \\ & \qquad
		\times
		\left[   \sin^2 \f{\pi y}{L_{{\it eff}}} \cosh^2 \f{\pi \tau}{L_{{\it eff}}} + \big( \sinh\f{\pi \alpha}{L_{{\it eff}}}  +\cos \f{\pi y}{L_{{\it eff}}} \sinh \f{\pi \tau}{L_{{\it eff}}}  \big) ^2\right],
		\\
		B(\tau,y) 
		  & = \sinh^2 \f{2\pi\alpha}{L_{{\it eff}}}                                                                                    
		+  4 \cos \f{2\pi y}{L_{{\it eff}}} (  \cos \f{2\pi y}{L_{{\it eff}}} - \cosh\f{2\pi\alpha}{L_{{\it eff}}} \cosh\f{2\pi\tau}{L_{{\it eff}}} ) \notag \\
		  & \quad   + 8 \cosh \f{2\pi \tau}{L_{{\it eff}}}\s{(\sinh^2\f{\pi\alpha}{L_{{\it eff}}} - \cosh\f{2\pi\alpha}{L_{{\it eff}}} 
		\sin ^2 \f{\pi y}{L_{{\it eff}}} - \sinh^2 \f{\pi \tau}{L_{{\it eff}}} )^2 + \f{1}{4} \sinh^2 \f{2\pi \alpha}{L_{{\it eff}}} \sin^2 \f{2\pi y}{L_{{\it eff}}}}.
	\end{align}
\end{widetext}

The EOW brane profile calculated from \eqref{eq:EOWprofileCoord} is shown in 
Fig.\ \ref{fig:EOWprofile2}.
Right at the moment of the quench, the EOW is sharply peaked at $y=0$,
and almost ``touches'' the boundary. 
For $t>0$, the peak splits into left- and right-moving ones.
They propagate away from the $y=0$.
These behaviors are consistent 
with the time-dependence of the stress-energy tensor.
For later times, the EOW profile looks more complicated.

\begin{figure}[t]
 \begin{center}
  \includegraphics[width=7.5cm]{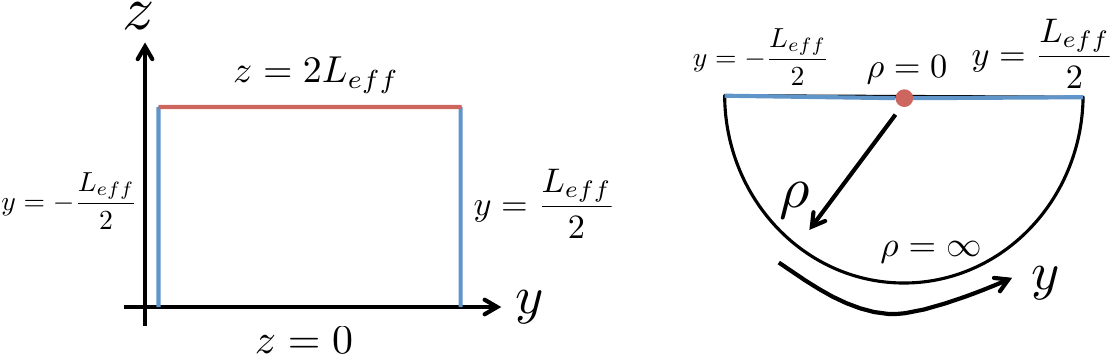}
 \end{center}
 \caption{The configuration of the tensionless EOW brane in the global AdS$_3$.
  The left panel is in the $(\eta,\tau,y)$ coordinate whereas the right panel is in the $(\rho,\tau,y)$ coordinate.
  In both cases, the $\tau$ directions are suppressed.}
 \label{fig:EOWglobal}
\end{figure}

We note that in the original Poincare coordinate \eqref{poincare},
the EOW intersects with the asymptotic boundary.
This does not appear to be the case in Fig.\ \ref{fig:EOWprofile2}.
As $y=\pm L_{{\it eff}}/2$ is the physical boundaries,
we may expect that the EOW intersects with the asymptotic boundary at these points.
The reason for this may be that our coordinates are ``not good.''
It is useful to illustrate what happens in the example of pure global AdS$_3$.
This corresponds to the $\alpha \to \infty$ limit
where we do not have any excitation. 
The relevant conformal transformation is obtained
by the $\alpha \to \infty$ limit of \eqref{eq:LocalQfinMap}, which leads to 
\be
f(w) = i e^{\f{i \pi}{L_{{\it eff}}}w}. 
\label{eq:GlobalT}
\ee 
The stress tensor \eqref{eq:SchwarzD} becomes
$\mathcal{L} = \bar{\mathcal{L}} = -\f{1}{4L_{{\it eff}}^2}$.
From this, the metric \eqref{eq:BanadosM} becomes 
\begin{align}
  ds^2 = \f{d z^2  + (1 + \f{z^2}{4L_{{\it eff}}^2})^2 d\tau^2
  +(1 - \f{z^2}{4L_{{\it eff}}^2})^2 dy^2 }{z^2 }.
\end{align}
This is actually the global AdS$_3$ metric. 
Changing the coordinate $\f{z}{2L_{{\it eff}}}= e^{-\rho}$
makes it easy to see the equivalence to the global AdS$_3$.
In this coordinate system, the metric is 
\begin{align}
  ds^2 = d\rho^2 + \cosh ^2\rho
  \left(
     \frac{d\tau}{ L_{{\it eff}} }
  \right)^2
  + \sinh ^2 \rho
   \left(\frac{dy}{L_{{\it eff}} }
  \right)^2.
\end{align}
On the other hand, using the map \eqref{eq:GlobalT} in \eqref{eq:BraneConfig} the brane profile becomes 
\be
z = 2L_{ {\it eff}}.
\ee
This actually corresponds to $\rho = 0$, which is just a point in a constant $\tau $ slice.
On the other hand, in the global AdS$_3$ case,
the tensionless EOW profile is known to be
given by the solution of
\cite{Numasawa:2018grg}
\be
\sinh \rho \sin \f{\pi y}{L_{ {\it eff}}} = 0.
\ee
Therefore, the EOW brane is located at $\rho = 0$ and also
at $y = \pm \f{L_{{\it eff}}}{2}$,
as depicted in Fig.\ \ref{fig:EOWglobal}.
However, we are missing
the $y = \pm \f{L_{{\it eff}}}{2}$ part in the formula \eqref{eq:BraneConfig}.

What we expect for the local quench is essentially the same.
The counterpart of $z = 2L_{{\it eff}}$ is captured by \eqref{eq:EOWprofileCoord} 
though they are generically a codimension one object rather than a point in $(z, y)$ plane, which is codimension two.
We expect that we are missing the counterpart of
$y = \pm \f{L_{{\it eff}}}{2}$ in the brane motion in the local quench problem.
We note that this missing problem does not occur 
in the local quench on an infinite line \cite{Caputa:2019avh}
where the EOW brane intersects with the AdS boundary at infinity.
It is interesting to comprehensively understand 
when this problem occurs and how to remedy it,
but we leave it to a future problem.

Note that the dynamics of the EOW brane in Fig.\ \ref{fig:EOWprofile2}
looks similar to the entanglement dynamics in Fig.\ \ref{fig:EOWprofileEnt}.
Through the holographic entanglement entropy formula
\cite{Ryu:2006bv,Ryu:2006ef,Hubeny:2007xt},
the entanglement entropy measures the distance between AdS boundary and the EOW brane.
The qualitative resemblance reflects this connection between spacetime geometry and entanglement.

\begin{figure*}[t]
\begin{center}
\includegraphics[width=6cm]{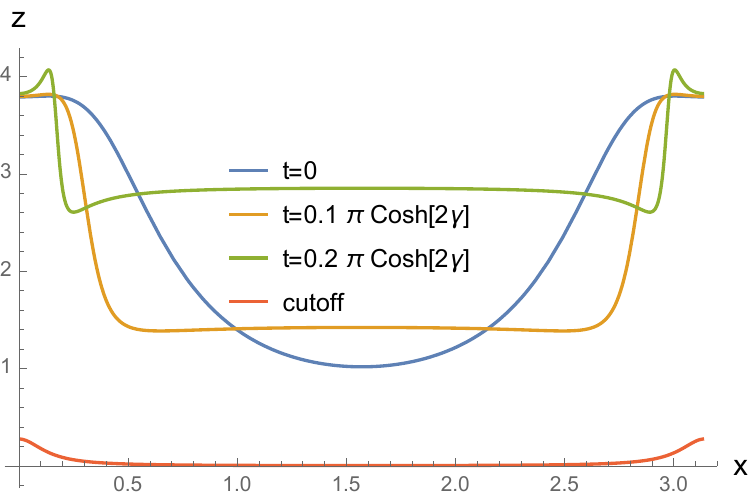}
\hspace{0.5cm}
\includegraphics[width=6cm]{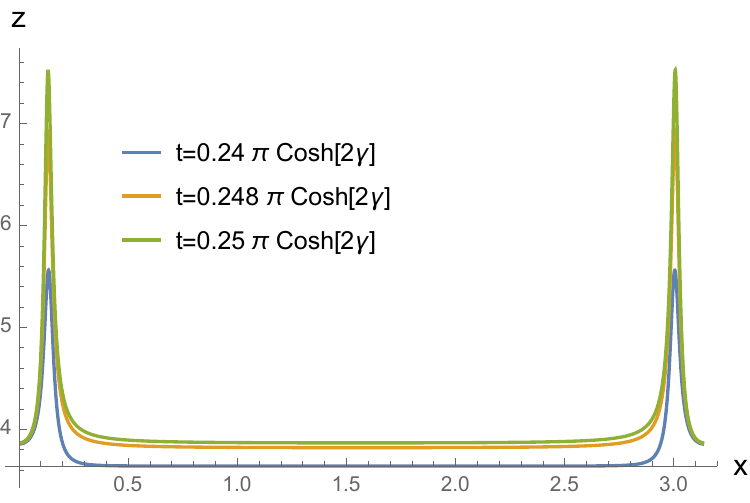}
\end{center}
\caption{ The EOW brane profile calculated from \eqref{eq:EOWprofileCoord}.
We put $L = \pi$ and $\gamma = 1$.
(Left) Early time behavior around $0 < t < 0.2  \pi \f{\pi}{L}\cosh(2\gamma)$.
The cutoff is taken to be $z = 0.01$.
(Right) The behavior near $t = \f{\pi}{4} \f{\pi}{L} \cosh(2\gamma)$. }
\label{fig:EOWprofile4}
\end{figure*}

\subsection{Holographic dual of M\"obius quench}

Now we can construct the holographic dual of the M\"obius quench
since we know the map from M\"obius quench to the local quench on a finite strip,
and also the holographic dual of the latter.
The dual geometry is given by
\ba
&ds^2  = 
\f{dz^2 -dt^2 + f_{\gamma}(x)^{-2} dx^2}{z^2}  \notag \\
& \qquad  + \hat{T}_{tt}dt^2 + 2 \hat{T}_{tx}dt dx + \hat{T}_{xx}dx^2 \notag \\
 &\qquad  + 4 z^2 \Big[f_{\gamma}(x)^{-2} \hat{T}_{tt} ^2  - \hat{T}_{tx}^2 \Big]( dx ^2 -  f_{\gamma}(x)^2dt^2 ), 
\ea
where the stress tensor profile is given by \eqref{eq:EMMobius}.
Now the cutoff is given by \eqref{eq:cutoff}, which is position dependent.
This position-dependent cutoff reproduces
the part of stress tensor \eqref{eq:WeylFactor} that comes from the Weyl transformation.
Because the dual geometry is given
by the dual geometry of the local quench with $\alpha$ in \eqref{eq:LocalqMobiusqRel}, 
we can use the EOW profile \eqref{eq:EOWprofileCoord}.
By changing the coordinate from $y$ to $x$ using the diffeomorphism \eqref{eq:LocalqMobiusqRel}, 
we obtain the EOW profile for the dual of M\"obius quenches.
The result is given by
\be
z(x,\tau)^2 = 
 \f{4L_{{\it eff}}^2}{\pi^2}  \f{\mathcal{A}(\tau,y)}{\mathcal{B}(\tau,y)} \label{eq:EOWprofileCoordMob},
\ee
where the numerator $\mathcal{A}(\tau,y)$
and the denominator $\mathcal{B}(\tau,y)$ are given by 
\begin{widetext}
\begin{align}
  &\mathcal{A}(\tau,x)
    =  \f{1}{f_{\gamma}(x)^2 \tanh ^2 (2\gamma)} \bigg[ \Big(2 \tanh (2\gamma) f_{\gamma}(x) \sinh ^2 \f{\pi \tau}{L_{eff}}
    + \f{1}{\cosh ^2 2\gamma} \Big)^2
    \notag \\
  &\qquad \qquad 
    - 4  f_{\gamma}(x) \tanh (2\gamma) \Big(1 -\cos \f{2\pi x}{L}\Big) \f{\sinh ^2 \f{\pi \tau}{L_{eff}}}{\cosh^22\gamma}\bigg],
  \\
&\mathcal{B}(\tau,x)
 =\f{1}{f_{\gamma}(x)^2 \tanh ^2 (2\gamma)} \Bigg[ 
\f{f_{\gamma}(x)^2}{\cosh^2 2\gamma} - 4 \tanh (2\gamma) g_{\gamma}(x) \Big( \f{1}{\cosh^2 2\gamma} + 2 f_{\gamma}(x) \sinh^2  \f{\pi \tau}{L_{eff}}  \Big) \notag \\
 & \qquad \qquad   + 4 \tanh (2\gamma) f_{\gamma}(x) \cosh \f{2\pi \tau}{L_{eff}} \s{ \big( \f{\cos \f{2\pi x}{L}}{\cosh ^2 2\gamma} + 2 f_{\gamma}(x)\tanh (2\gamma) \sinh ^2 \f{\pi \tau}{L_{eff}}\big)^2  + \f{\sin^2 \f{2\pi x}{L}}{\cosh ^4 2\gamma}}
   \Bigg].
   \label{eq:EOWprofileMob}
\end{align}
Here we recall $g_{\gamma}(x) = \tanh 2\gamma - \cos \f{2\pi x}{L}$ and $f_{\gamma}(x) =  1 -\tanh 2\gamma \cos \f{2\pi x}{L}$.
After analytically continuing to the Lorentzian time $\tau \to it$,
we obtain the time dependence of the EOW profile.
\end{widetext}

The EOW profile is shown in Fig.\ \ref{fig:EOWprofile4}.
Note that we again encounter the same problem as in the case of the local quench
where we cannot follow the dynamics of the EOW profile between $t \in [L_{{\it eff}}/4,3L_{{\it eff}}/4]$.
From the expression \eqref{eq:EOWprofileMob},
we find that there is a periodicity in real time with the period $L_{{\it eff}} = L \cosh 2\gamma$.

\section{Conclusion} \label{sec:conclusion}

In this paper, we studied the dynamics after
the local quench on finite intervals and
the M\"obius quench in (1+1)d CFT.
First, we found that the M\"obius quench can be obtained
from the local quench by diffeomorphism and Weyl transformations.
In the holographic setups, we employ the AdS/BCFT correspondence
and study the motion of the EOW brane.
We also compare this brane motion with
the entanglement dynamics for an interval.
The brane dynamics qualitatively agrees with the entanglement dynamics.

There is in principle a vast class of quantum quench problems
that we can consider in (1+1)d CFT. 
As demonstrated here, we expect that some of them are related to each other.
It would be interesting to explore
this type of equivalence relation further
beyond the specific examples considered in this paper.
This may lead to a classification of possible dynamical behaviors using the equivalence
relation.

\acknowledgements
SR is supported by the National Science Foundation under 
Award No.\ DMR-2001181, and by a Simons Investigator Grant from
the Simons Foundation (Award No.~566116).
TN is supported by MEXT KAKENHI Grant-in-Aid for Transformative Research Areas A ``Extreme Universe'' (22H05248) and JSPS KAKENHI Grant-in-Aid for Early-Career Scientists (23K13094).
MN is supported by funds from University of Chinese Academy of Sciences (UCAS), and funds from the Kavli Institute for Theoretical Sciences (KITS).
MT is supported by an appointment to the YST Program at the APCTP through the Science and Technology Promotion Fund and Lottery Fund of the Korean Government. MT is also supported by the Korean Local Governments -
Gyeongsangbuk-do Province and Pohang City. JKF is supported by the Institute for Advanced Study and the National Science Foundation under Grant No.~PHY-2207584.
This work is supported by the Gordon and Betty Moore Foundation 
through Grant GBMF8685 toward the Princeton theory program.

\bibliography{Mobiusbunken.bib}

\end{document}